\documentclass[10pt,letterpaper,compsoc,conference]{iiswc23}

\usepackage{cite}
\usepackage{amsmath,amssymb,amsfonts}
\usepackage{algorithmic}
\usepackage{graphicx}
\usepackage[dvipsnames]{xcolor}
\usepackage[final]{microtype}
\usepackage[italic]{mathastext}
\usepackage{libertine}
\usepackage[T1]{fontenc}
\usepackage{textcomp}
\usepackage[varqu,varl]{zi4}
\usepackage[all]{nowidow}
\usepackage[auth-lg,affil-it]{authblk}
\usepackage[keeplastbox]{flushend}
\usepackage[left=1.8cm, right=1.8cm, top=2.3cm, bottom=2.35cm]{geometry}
\usepackage{tikz}

\usepackage[subtle]{savetrees}
\newcommand\copyrighttext{%
  \footnotesize \textcopyright 2023 IEEE. Personal use of this material is permitted.
  Permission from IEEE must be obtained for all other uses, in any current or future
  media, including reprinting/republishing this material for advertising or promotional
  purposes, creating new collective works, for resale or redistribution to servers or
  lists, or reuse of any copyrighted component of this work in other works.
  DOI: paper DOI.}
\newcommand\copyrightnotice{%
\begin{tikzpicture}[remember picture,overlay]
\node[anchor=south,yshift=10pt] at (current page.south) {\fbox{\parbox{\dimexpr\textwidth-\fboxsep-\fboxrule\relax}{\copyrighttext}}};
\end{tikzpicture}%
}

\begin{document}


\title{Optimising GPGPU Execution Through Runtime Micro-Architecture Parameter Analysis}


\renewcommand\Authsep{\quad}
\renewcommand\Authand{\quad}
\renewcommand\Authands{\quad}



\author[1,2]{Giuseppe M. Sarda}
\author[1]{Nimish Shah}
\author[2]{Debjyoti Bhattacharjee}
\author[2]{Peter Debacker}
\author[1,2]{Marian Verhelst}
\affil[1]{KU Leuven, Leuven, Belgium}
\affil[2]{imec, Leuven, Belgium }

\maketitle


\begin{abstract}
GPGPU execution analysis has always been tied to closed-source, proprietary benchmarking tools that provide high-level, non-exhaustive, and/or statistical information, preventing a thorough understanding of bottlenecks and optimization possibilities. Open-source hardware platforms offer opportunities to overcome such limits and co-optimize the full {hardware-mapping-algorithm} compute stack. Yet, so far, this has remained under-explored. 
In this work, we exploit micro-architecture parameter analysis to develop a hardware-aware, runtime mapping technique for OpenCL kernels on the open Vortex RISC-V GPGPU. Our method is based on trace observations and targets optimal hardware resource utilization to achieve superior performance and flexibility compared to hardware-agnostic mapping approaches.
The technique was validated on different architectural GPU configurations across several OpenCL kernels.
Overall, our approach significantly enhances the performance of the open-source Vortex GPGPU,
contributing to unlocking its potential and usability.
\end{abstract}

\section{Introduction}
\copyrightnotice
\begin{figure}[!ht]

    \centering
    \includegraphics[width=\linewidth]{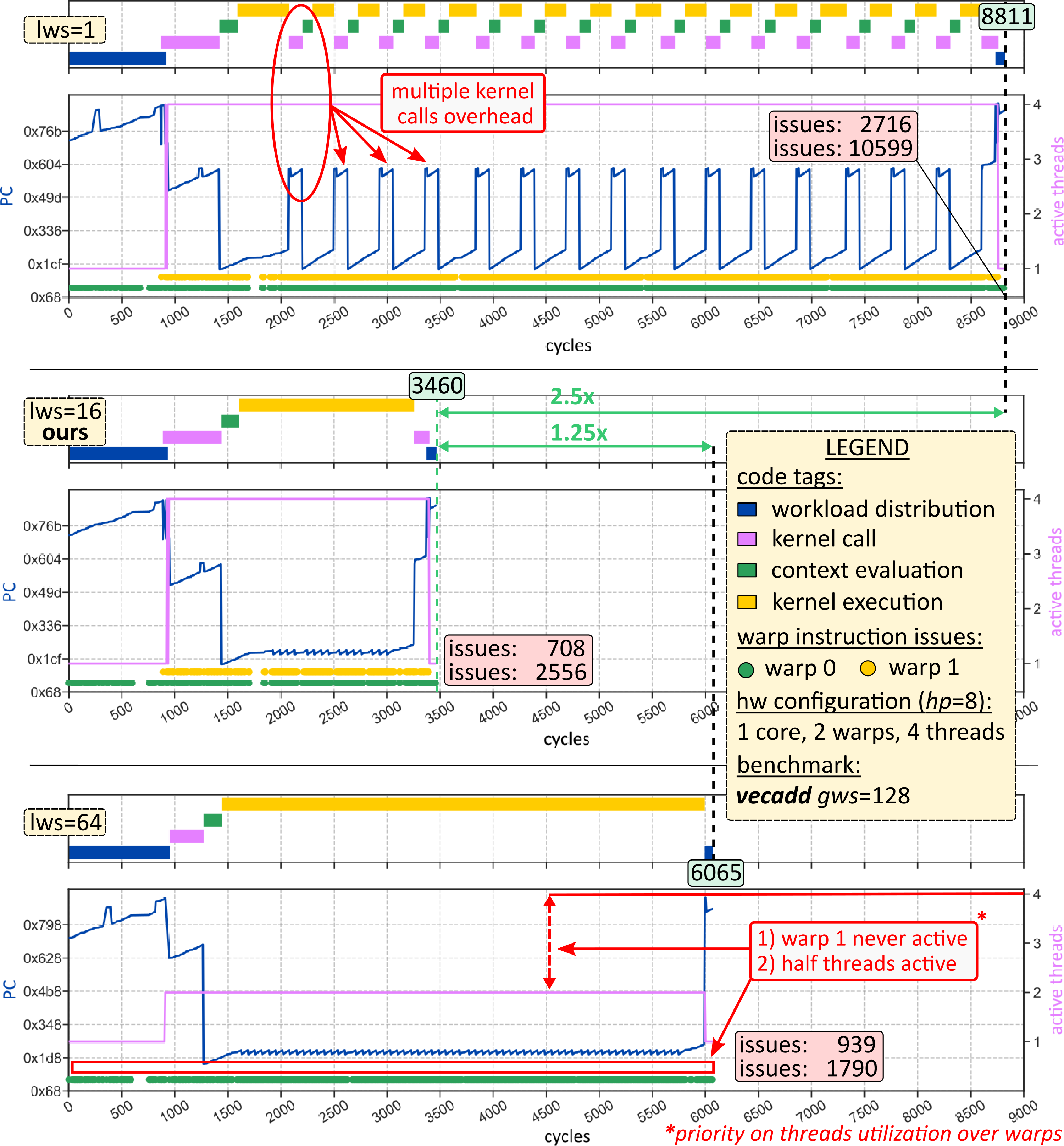}
    \caption{Execution traces of the \textit{vecadd} kernel under 4 different \textit{lws}. Each plot shows tagged instruction wavefronts, the PC, the active thread mask and the timestamp of instruction issues from different warps.}
    \label{fig:vecadd}
    \vspace{-2ex}
\end{figure}
\begin{figure*}
    \centering
    \includegraphics[width=\linewidth]{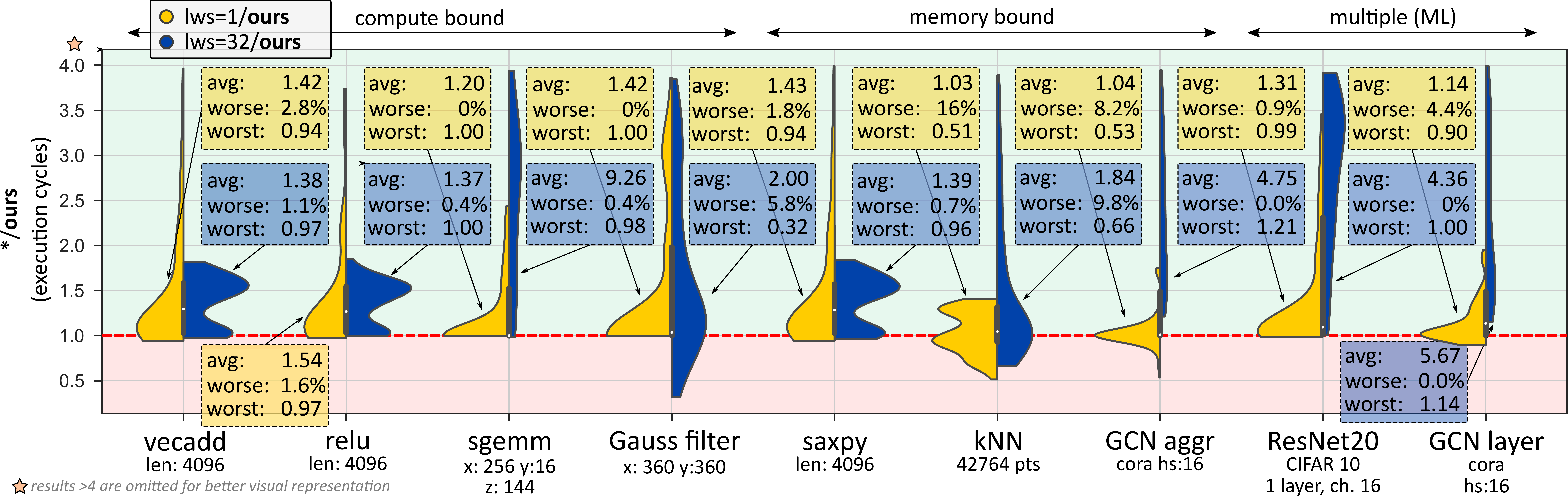}
    \caption{Violin plots showing the comparison (ratio) of latencies from our methodology vs fixed (\textit{lws}=32, right in blue) and naive mapping (\textit{lws}=1, left in yellow) on 450 different HW architectural configurations. Data tables show the average, the worst result, and the result count <1 (x/450) in percentage.} 
    \label{fig:validation}
    \vspace{-2ex}
\end{figure*}
The rise of AI algorithms, demanding more compute power, and the slowdown of Moore's law, hindering performance improvements from pure silicon technology scaling, led to the development of data-parallel, application-specific architectures like GPUs and NPUs. However, most architectures are proprietary, facing challenges regarding versatility and closed-source limitations.
An opportunity to address these issues lies in open-source hardware and software platforms, with the Vortex RISCV-based GPGPU \cite{vortex} being a promising and versatile option for exploration and characterization across various algorithms.
The open hardware, ISA, and software stack allow, in fact, deep analysis and understanding of the execution in the platform. This enables recognizing the bottlenecks and more chances to co-optimize the GPU across the whole stack.
Our work demonstrates the impact of leveraging low-level, micro-architecture information to improve, with a single flexible approach, the execution of several kernels on this open-source Vortex GPU. To ensure algorithmic versatility, we chose different math kernels and synthetic layers from typical Deep Neural Networks \label{DNN} (DNN) \cite{resnet} and Graph Convolutional Networks \label{GCN} (GCN) \cite{GCNp}. 

Specifically, this paper reports our contributions in terms of: 

\vspace{0.5ex}
\noindent $\bullet$ trace analysis from the RISC-V-based Vortex GPGPU

\vspace{0.5ex}
\noindent $\bullet$ our hardware-aware optimal, runtime OpenCL kernel mapping

\vspace{0.5ex}
\noindent $\bullet$ the impact of our technique on the execution of typical math kernels and layers in DNN and GCN

\section{Analysis and workload mapping}

The Vortex POCL compiler \cite{vpocl} accepts standard OpenCL kernels. Compiled code is linked with runtime libraries, which take care of initializing the platform and spatially and temporally mapping the parallel instances of the kernel.


Before calling the kernel, the Vortex runtime maps the workload equally across cores. Within each core, the kernel
iterations are further distributed among threads first and then warps, depending on the \textit{$local\_work\_size$} (\textit{lws}). This \textit{lws} is one of the arguments passed by the host platform when calling the GPGPU execution \cite{opencl} and, in essence, determines the iterations each thread loops around the kernel for each internal call.
Fig. \ref{fig:vecadd} shows an example of the impact of changing the \textit{lws} parameter on the execution of 128-element vector addition (\textit{vecadd}) in a simple 1 core, 2 warps, 4 threads (\textit{1c2w4t}) GPU configuration; the plot provides the PC, the instruction thread mask, and warp issue information over time, for 4 different \textit{lws} values. For better visualization, we tagged instruction addresses with different semantic sections of the code (shown as a waveform graph above every plot).
Depending on the relationship between the \textit{lws} mapping parameter, the global workload size \textit{gws} (e.g., the total iterations the kernel will be executed), and the hardware parallelism \textit{hp}, resolved in Eq. \ref{eq:lws}, there are 3 possible scenarios (\textit{gws}=128 and \textit{hp}=8 in the example): 

\vspace{0.5ex}
\noindent $\bullet$ \textbf{\textit{lws} $<$ \textit{gws}/\textit{hp}}: the software will spawn more warps than the hw can support. The execution will be scheduled at different timesteps with multiple kernel calls, cfr. the uppermost "lws=1" scenario in Fig. \ref{fig:vecadd}.

\vspace{0.5ex}
\noindent $\bullet$ \textbf{\textit{lws} = \textit{gws}/\textit{hp}}: all warps will be loaded in parallel into the hardware with a single kernel call, cfr. the "lws=16" scenario.

\vspace{0.5ex}
\noindent $\bullet$ \textbf{\textit{lws} $>$ \textit{gws}/\textit{hp}}: all warps will be loaded in parallel into the hardware, yet with reduced hardware utilization, cfr. "lws=32/64" scenarios.
\vspace{0.5ex}

The optimal \textit{lws} value is, hence, both hardware and algorithm dependent, and can be determined as:
\begin{equation}
    lws = \frac{gws}{hp} \text{, with }hp =  cores \times warps \times threads
    \label{eq:lws}
\end{equation}
This value can be evaluated at runtime based on the hardware properties and the workload size, without being explicitly specified by the programmer.


\section{Validation}
To validate our observations, we analyzed the execution on 450 different hardware GPU configurations, spanning from 1 core, 2 warps, and 2 threads (\textit{1c2w2t}) to \textit{64c32w32t}, running stand-alone math kernels and combined ones for DNN and GCN layers. We compared our mapping, obtained with Eq. \ref{eq:lws}, with a naive (\textit{lws}=1) (e.g., never unrolling the kernel temporally over one thread) and a fixed one (\textit{lws}=32). \\
Fig. \ref{fig:validation} compares the resulting number of execution cycles, plotting the ratio between other mappings and ours.
Our technique shows an average $1.3\times$ and $3.7\times$ performance boost for the math kernels over the \textit{lws}=1 mapping and the \textit{lws}=32, respectively. 

From the plots, we can observe that, across different hw solutions, providing the kernel execution with the same \textit{lws} results in a large performance variability: from optimal to up to 20x slower. This proves that our mapping, with hw and sw awareness, can adapt and benefit a wide range of kernels. 
Note that in a few specific hw configurations, spawning more or less warps can bring small benefits to the execution (because of e.g., reduced overhead, improved memory bandwidth utilization, etc). This is visible in the plot, as some distribution cut-offs are slightly below the bold, red line on 1. Also, when the hardware parallelism \textit{hp} exceeds the \textit{gws} of the executed kernel, Eq. \ref{eq:lws} resolves to \textit{lws=1}. This justifies the peaks around 0 on the left, yellow side of the violin plots.
Finally, the Gaussian blur filter, the near-neighbor search, and GCN aggregation kernels show atypical trends, we will explore the reasons in future work.

\section{Conclusions}
In this work, we analyzed the software-to-hardware mapping flow on Vortex through execution traces, showing a method to runtime optimize the \textit{lws} parameter and abstract its hardware impact to the programmer. We validated the approach on the diverse math kernels and ML layers.
It is clear that other factors still impact the runtime kernel execution in Vortex. Going further, these will be analyzed in more depth, to improve the end-to-end execution of neural networks from a combined software and hardware point of view.

\section*{Acknowledgments}
Project funded by the European Research Council (ERC) grant No. 101088865, EU H2020 grant No. 101070374, the Flanders AI Research Program, and the KU Leuven.

\bibliographystyle{IEEEtranS}
\bibliography{main.bib}

\end{document}